\documentclass[10pt,letterpaper]{article}
\usepackage{opex3}
\usepackage{amsmath}
\usepackage{upgreek}

\begin{document}

\title{Optical Quadratic Measure Eigenmodes}
\author{M. Mazilu, J. Baumgartl, S. Kosmeier, and K. Dholakia}
\address{SUPA, School of Physics and Astronomy, University of St Andrews, North Haugh, Fife, KY16~9SS, UK}

\email{michael.mazilu@st-andrews.ac.uk} 

\begin{abstract}
We report a mathematically rigorous technique which facilitates the optimization of various optical  properties of electromagnetic fields. The technique exploits the linearity of electromagnetic fields along with the quadratic nature of their interaction with matter. In this manner we may decompose the respective fields into optical quadratic measure eigenmodes (QME). Key applications include the optimization of the size of a focused spot, the transmission through photonic devices, and the structured illumination of photonic and plasmonic structures. We verify the validity of the QME approach through a particular experimental realization where the size of a focused optical field is minimized using a superposition of Bessel beams.
\end{abstract}


\section{Introduction}

The decomposition of fields into eigenmodes is a well established technique to solve various problems within physical sciences. The most prominent example refers to Schr\"{o}dinger's equation, within the field of quantum mechanics, where energy spectra of atoms are determined via the eigenvalue spectra and associate wavefunctions of the Hamiltonian operator. Indeed, electron orbits are eigenmodes of the energy, angular momentum, and spin operators~\cite{cohen-tannoudji1977} and as such they deliver fundamental insights into the physics of atoms. Within classical mechanics, modes of vibration of music instruments give, for example, their resonant frequencies while their spectrum is associated with the shape of the instrument~\cite{Kac:1966p9889}. In the optical domain, mode decomposition is used in order to describe light propagation within waveguides~\cite{Sudbo:1993p9727}, photonic crystals~\cite{Bienstman:2001p9890}, and optical cavities~\cite{Reithmaier:1997p9888}. In the case of waveguides and photonic crystals, for example, eigenmodes describe electromagnetic fields that are invariant in their intensity profile as they propagate along the fibre or crystal. Additionally, these modes are orthogonal and as such light coupled to one of these modes remains, in theory, in this mode forever. This optical mode decomposition can be expanded to include additional operators such as orbital and spin angular momentum~\cite{Mazilu:2009p9267}.

In this paper, we report a novel method which we term ``Quadratic Measure Eigenmodes (QME)'' which represents a generalization of the powerful concept of eigenmode decomposition within the field of optics. Crucially, it is shown that eigenmode decomposition is applicable to the case of any quadratic measure which is defined as a function of the electromagnetic field. Prominent examples of optical quadratic measures include the energy density and the energy flux of electromagnetic fields. The QME method makes it possible to describe an optical system and its response to incident electromagnetic fields as a simple mode coupling problem and to determine the optimal ``excitation'' for the given measure considered. Intuitively, a superposition of initial fields is optimized in a manner that the minimum/maximum measure is achieved. For instance, the transmission through a pinhole is optimized by maximizing the energy flux through the pinhole. 

From a theoretical perspective, the QME optimization method is mathematically rigorous and may be distinguished  from the multiple techniques currently employed ranging from genetic algorithms~\cite{Assion:1998p9709} and random search methods~\cite{Dennis:2010p9723} to direct search methods~\cite{Thomson:2008p8158}. The major challenge encountered in any such approximate optimization and engineering of optical properties is the fact that electromagnetic waves interfere. As such the interference pattern not only makes the search for an optimum beam problematic but crucially renders the superposition found unreliable, as the different algorithms may converge on different local minima which are unstable with respect to the different initial parameters in the problem. In contrast, our proposed QME method yields a unique solution to the problem and directly determines the optimum (maximal/minimal) measure possible.

In the first part of the paper, we introduce the background of the QME theory and show its properties in a general context of optimizing the quadratic measures of interfering waves. In the second part, we apply the QME formalism to maximize the transmission through apertures and to minimize the focal spot size. For these applications, we describe, respectively, the electromagnetic field as a superposition of scalar Laguerre-Gaussian beams and vectorial Bessel beams. In the final part of the paper, we report a particular experimental implementation of the QME method using computer controlled spatial light modulators to squeeze the spot size of a superposition of Bessel beams. The paper concludes with a discussion of the particular results obtained and with general comments on the versatility of the QME method to a wide range of problems.

\section{Fundamental concepts} 
\label{s:fundcon}

The QME method is based upon two fundamental properties of the electromagnetic field and its interactions. Firstly, the approach relies on the linearity of the electromagnetic fields, \emph{i.e.}, the sum of two solutions of Maxwell's equations is itself a solution of them. As we consider free space propagation, this criteria is satisfied. The second property relates to the interaction of the electromagnetic field with its environment. All such interactions can be written in the form of quadratic expressions with respect to the electric and magnetic fields. Examples include the energy density, the energy flow, and Maxwell's stress tensor. This allows us to designate appropriate QME to various parameters (\emph{e.g.} spot size) and subsequently ascertain the optimal eigenvalue which, in the case of a spot size operator, yields a sub-diffraction optical spot. In this section, we present the details of the theory underpinning our approach.

\subsection{Theoretical background}
\label{Sec: TheoreticalBackground}

\paragraph{Electromagnetic waves}

To demonstrate our method, we consider monochromatic solutions of the free space Maxwell's equations:
\begin{eqnarray}\label{Eq:Maxwell}
\nabla\cdot\epsilon_{0}\mathbf{E}  =  0, \;\;\;\; 
\nabla\cdot\mu_{0}\mathbf{H}  =  0, \;\;\;\;
 \nabla\times\mathbf{E}  =  -\text{i}\mu_{0} \omega \mathbf{H}, \;\;\;\;
 \nabla\times\mathbf{H}  =  \text{i} \epsilon_{0}\omega\mathbf{E},
\end{eqnarray}
where $\mathbf{E}$ and $\mathbf{H}$ are the spatial part of the electric and magnetic 
vector fields and where $\epsilon_{0}$ and $\mu_{0}$ denote the vacuum permittivity and permeability. The time dependent carrier wave is given by $\exp(\text{i}\omega t)$. These monochromatic solutions of Maxwell's equations can be written in an integral form linking the electromagnetic fields on the surface $A$ with the fields at any position $\mathbf r$,
\begin{equation}
\label{diffract}
\mathcal{F}_i(\mathbf r) =\int_{A}\mathbf{P}_{ij}(\mathbf{r},\mathbf{r}')\mathcal{F}_j(\mathbf{r}')dS'
\end{equation}
where $\sqrt{2} \mathcal F=(\sqrt{\epsilon_0}\mathbf E,\sqrt{\mu_0}\mathbf H)$ is a shorthand for the two electromagnetic fields having six scalar components $\mathcal F_i$. The integration kernel $\mathbf P_{ij}$ corresponds to a propagation operator giving rise to different vector diffraction integrals such as  Huygens, Kirchhoff~\cite{KRAUS:1990p7671}, and Stratton-Chu~\cite{Stratton:1939p2484}.

\paragraph{Quadratic measures}

Crucially all ``linear'' and measurable properties of the electromagnetic field can be expressed as quadratic forms of the local vector fields and are therefore termed \emph{quadratic measures}. For instance, the time averaged energy density of the field is proportional to $\mathcal{F}^{\ast}\cdot\mathcal{F}=1/2(\epsilon_0\mathbf{E}^{\ast}\cdot\mathbf{E}+\mu_0\mathbf{H}^{\ast}\cdot\mathbf{H})$ while the energy flux to $1/2 (\mathbf{E}^{\ast}\times\mathbf{H}+\mathbf{E}\times\mathbf{H}^{\ast})$. The asterisk $^{\ast}$ stands for the complex conjugate. Integrating the first quantity over a volume determines the total electromagnetic energy in this volume, and integrating the normal energy flux across a surface yields the intensity of the light field incident on this surface. All the quadratic measures $m_{\kappa}$ can be represented in a compact way by considering the integral
\begin{equation}
\label{QM:genl}
m_\kappa=\int_V\mathcal{F}_i^{\ast}\boldsymbol{\kappa}_{ij}\mathcal{F}_j d \mathbf{r} =\langle\mathcal{F}|\boldsymbol{\kappa}|\mathcal{F}\rangle _V 
\end{equation}
where the kernel $ \kappa_{ij} = \kappa^{\dagger}_{ji} $ is Hermitian with $\dagger$ the adjoint operator including boundary effects for finite volumes. Table \ref{tab} enumerates some operators associated to common quadratic measures. The integrand part of most of these quadratic measures corresponds to the conserving densities, which together with the associated currents are Lorentz invariant~\cite{Mazilu:2009p9267}. The volume, over which the integral is taken, does not need to be the whole space and can be a region of space, a surface, a curve, or simply multiple points. To account for this general integration volume, we broadly term it the region of interest (ROI) in the following. 

\begin{table}[htb]
\begin{center}\begin{tabular}{|c|c|}
\hline
Operator  & $2\mathcal F_i^{\ast} \boldsymbol \kappa_{ij}  \mathcal F_j $ \\ \hline
EO & $  \epsilon_0 \mathbf{E}^*\cdot\mathbf{E}+\mu_0 \mathbf{H}^*\cdot\mathbf{H}$ \\
IO  &$( \mathbf{E}^*\times\mathbf{H}+\mathbf{E}\times\mathbf{H}^*) \cdot \mathbf{{e}}_k$ \\
SSO  & $\mathbf r^2  ( \mathbf{E}^*\times\mathbf{H}+\mathbf{E}\times\mathbf{H}^*) \cdot \mathbf{{e}}_k$ \\
LMO  & $  \epsilon_0 \mathbf{E}^*\cdot(i\partial_k)\mathbf{E}+\mu_0 \mathbf{H}^*\cdot(i\partial_k)\mathbf{H}  $ \\
OAMO  & $ \epsilon_0 \mathbf{E}^*\cdot(i\mathbf r\times \nabla)_k\mathbf{E}+\mu_0 \mathbf{H}^*\cdot (i\mathbf r\times \nabla)_k\mathbf{H} $  \\
CSO  &$i( \mathbf{E}^*\cdot\mathbf{H}-\mathbf{H}^*\cdot\mathbf{E})$ \\
\hline
\end{tabular}
\caption{Common quadratic measure operators including the energy operator (EO), intensity operator (IO), spot size operator (SSO), linear momentum operator (LMO), orbital angular momentum operator (OAMO), and circular spin operator (CSO). The vector operators include the subscript $k$ indicating the different coordinates and $\mathbf{e}_k$ the associated unit vectors.}
\label{tab}\end{center}
\end{table}

\paragraph{Quadratic measures eigenmodes} Finally, using the general definition (\ref{QM:genl}) of the quadratic measure it is possible to define a Hilbert sub-space, over the solutions of Maxwell's equations, with the energy operator  defining the inner product. Furthermore, any general quadratic measure defined by (\ref{QM:genl}) can be represented in this Hilbert space by means of its spectrum of eigenvalues and eigenfunctions defined by:
 \begin{equation}
 \label{QME}
\lambda \mathcal F_i(\mathbf r_2)=  \int_V  \kappa_{ij} (\mathbf  r_2, \mathbf r_1) \mathcal F_j(\mathbf r_1) d \mathbf r_1 = \boldsymbol \kappa | \mathcal F \rangle_V \nonumber. 
\end{equation}
Depending upon the kernel $\kappa_{ij}$ or operator $ \boldsymbol \kappa$, the eigenvalues $\lambda$ form a continuous or discrete real valued spectrum which can be ordered. This gives direct access to the solution of Maxwell's equations with the largest or smallest measure. The eigenfunctions are orthogonal to each other ensuring simultaneous linearity in both field and measure. 

In the following, we study the case of different quadratic measure operators and their spectral decomposition into the QME. The convention for operator labeling we adopt is to use the shorthand QME followed by a colon and a shorthand of the operator name. In the following, we discuss some examples of different quadratic measure operators.

\paragraph{QME: Intensity operator (QME:IO)}
The quadratic measure corresponding to the QME:IO measures the electromagnetic energy flow across a surface ROI: 
\begin{equation}
\label{eq:QMEIO2}
m^{(0)}=\frac{1}{2}\int_{\textrm{ROI}} ( \mathbf{E}^*\times\mathbf{H}+\mathbf{E}\times\mathbf{H}^*) \cdot \mathbf n dS
\end{equation}
where $\boldsymbol \kappa$ is chosen such that it corresponds to the projection of the Poynting vector on the normal $\mathbf n$ to the surface. The eigenvector decomposition of this operator can be used to maximize the optical throughput through a pinhole or to minimize the intensity in dark spots. Considering a closed surface ROI surrounding an absorbing particle, the QME approach gives access to the field that either maximizes or minimizes the absorption of this particle. 

\paragraph{QME: Spot size operator (QME:SSO)}
One way to define the spot size of a laser beam is by measuring the second order intensity moment (SOIM) $w$~\cite{Paschotta2008EncOfLasPhysAndTechn}. $w$ can be expressed as 
\begin{align}
\label{Eq: SOM}
w &= 2\sqrt{\frac{m^{(2)}}{m^{(0)}}},
\end{align}
with $m^{(0)}$ as defined in Eq.~\eqref{eq:QMEIO2} and the QME:SSO defined by
\begin{equation}
\label{eq:QMESSO2}
m^{(2)}=\frac{1}{2}\int_{\textrm{ROI}}  |\mathbf r-\mathbf r_0|^2  ( \mathbf{E}^*\times\mathbf{H}+\mathbf{E}\times\mathbf{H}^*) \cdot \mathbf n dS,
\end{equation}
where $\mathbf r$ is the position vector and $\mathbf r_0$ the centre of the beam. Accordingly, the square root of the QME:SSO eigenvalues multiplied by two gives direct access to the respective beam size provided that the intensity is normalized to one within the ROI, i.e., $m^{(0)}=1$;

\paragraph{QME: Optical force operator (QME:OFO)}
The optical force acting on a scattering object can be calculated by considering the momentum flux, given by Maxwell's stress tensor, across a surface ROI, surrounding the scattering particle. There are three quadratic measures associated with the optical force acting on the particle, one for each orthogonal direction:
\begin{eqnarray}
\mathbf{m}^{(F)} &=& \int_{ROI} \frac{\epsilon_0}{2} \mathbf E^* (\mathbf E \cdot\mathbf n) +\frac{\mu_0}{2}\mathbf H^*(\mathbf H \cdot\mathbf n)  -\frac{1}{4}(\epsilon_0\mathbf E^* \cdot \mathbf E +\mu_0\mathbf H^* \cdot \mathbf H )\mathbf n \;dS\label{mst}  
\end{eqnarray}
where $\mathbf{m}^{(F)}=\left(m_{x}^{(F)},m_{x}^{(F)},m_{x}^{(F)}\right)$ corresponds to the measure of the force. The eigenvector decomposition of this operator can be used to maximize the optical scattering and optical trapping force on microparticles. This approach can be extended through the use of the angular momentum operator, defined locally by $\textrm{i}\mathbf r\times \nabla$. 

\section{Applications}
\label{s:sec3}

In this section, we put the QME concept into practice and provide a couple of examples which demonstrate the striking applicability of the QME formalism to answer different questions within the field of optics. One such question is determining the largest intensity that may be transmitted through a given optical structure such as metallic apertures~\cite{GarciaVidal:2005p9724}. Another question is what is the smallest optical spot one may achieve.
Before we discuss these questions, we show  how the operator formalism presented in section~\ref{s:fundcon} is rendered into a practical formalism which we have also applied in the experiments described in section~\ref{s:sec4}. 

\subsection{Practical implementation of the QME concept}
\label{s:sec31}

For practical purposes, in particular in terms of the experimental realization of the QME concept, the optimization procedure is spatially separated; that is we consider 1) an initial plane at the propagation distance $z=z_{1}$ where we can superpose a set of $i=1\dots{}N$ fields $\mathbf{E}_{i}(x,y,z)$ and $\mathbf{H}_{i}(x,y,z)$ ($N>1$) shaped both in amplitude and phase and 2) a target plane at the propagation distance $z=z_{2}$ where we have the ROI within which the optimization is actually carried out. Due to linearity a superposition of fields in the initial plane is rendered into a superposition in the target plane both featuring the same set of superposition coefficients:
\begin{equation}
\label{eq:sec311}
\left.
\begin{array}{rcl}
\displaystyle
\mathbf{E}(x,y,z_{1})&=&
\displaystyle
\sum_{i=1}^{N}a_{i}\mathbf{E}_{i}(x,y,z_{1})\\
\displaystyle
\mathbf{H}(x,y,z_{1})&=&
\displaystyle
\sum_{i=1}^{N}a_{i}\mathbf{H}_{i}(x,y,z_{1})
\end{array}
\right\}
\longrightarrow
\left\{
\begin{array}{rcl}
\displaystyle
\mathbf{E}(x,y,z_{2})&=&
\displaystyle
\sum_{i=1}^{N}a_{i}\mathbf{E}_{i}(x,y,z_{2})\\
\displaystyle
\mathbf{H}(x,y,z_{2})&=&
\displaystyle
\sum_{i=1}^{N}a_{i}\mathbf{H}_{i}(x,y,z_{2})
\end{array}
\right.
.
\end{equation}
Based on this, the QME:IO defined in \eqref{eq:QMEIO2} can be rewritten as 
\begin{equation}
\label{eq:s312}
m^{(0)}=\mathbf{a}^{\ast}{\mathbf{M}}^{(0)}\mathbf{a}.
\end{equation}
The matrix $\mathbf{M}^{(0)}$ is a $N\times{}N$ matrix with the elements given by the overlap integrals 
\begin{equation}
\label{eq:s313}
M_{ij}^{(0)}=\frac{1}{2}\int_{\textrm{ROI}} \left(\mathbf{E}_{i}^{\ast}(x,y,z_{2})\times\mathbf{H}_{j}(x,y,z_{2})+\mathbf{E}_{i}(x,y,z_{2})\times\mathbf{H}_{j}^{\ast}(x,y,z_{2})\right) \cdot \mathbf{n} dS.
\end{equation}
This matrix is equivalent to the QME:IO on the Hilbert subspace defined by the fields $\left\{\mathbf{E}_{i}(x,y,z_{2}),\mathbf{H}_{i}(x,y,z_{2})\right\}$. $\mathbf{M}^{(0)}$ is Hermitian and positive-definite which implies that its eigenvalues $\lambda_{k}^{(0)}$ ($k=1\dots{}N$) are real and positive and the eigenvectors $\mathbf{v}_{k}^{(0)}$ are mutually orthogonal. Accordingly, the largest eigenvalue $\lambda_{\max}^{(0)}=\max\left(\lambda_{k}^{(0)}\right)$ and the associated eigenvector $\mathbf{v}_{\max}^{(0)}$ will deliver the initial plane ($z=z_{1}$) or target plane ($z=z_{2}$)  superposition
\begin{equation}
\label{eq:s314} \left\{\mathbf{E}_{\max}(x,y,z)=\sum_{i=1}^{N}v_{\max,i}^{(0)}\mathbf{E}_{i}(x,y,z),\ \mathbf{H}_{\max}(x,y,z)=\sum_{i=1}^{N}v_{\max,i}^{(0)}\mathbf{H}_{i}(x,y,z)\right\},
\end{equation}
which maximizes the intensity within the ROI. We remark, that the case of a superposition of scalar fields $u_{i}(x,y,z)$, the matrix operator \eqref{eq:s313} takes on the simpler form
\begin{equation}
\label{eq:s315}
M_{ij}^{(0)}=\int_{\textrm{ROI}} u_{i}^{\ast}(x,y,z_{2})u_{j}(x,y,z_{2})dS,
\end{equation} 
an approach used in section~\ref{s:sec32} to maximize the transmission through a pinhole.

Similar to the QME:IO, the QME:SSO defined in Eq.~\eqref{eq:QMESSO2} can be rewritten as
\begin{equation}
\label{eq:s316}
m^{(2)}=\mathbf{b}^{\ast}{\mathbf{M}}^{(2)}\mathbf{b},
\end{equation}
where ${\mathbf{M}}^{(2)}$ and $\mathbf{b}$ must be represented in the intensity normalised base
\begin{equation}
\label{eq:s317}
\left\{
\widetilde{\mathbf{E}}_{k}(x,y,z_{2})=\sum_{i=1}^{N}\frac{v_{k,i}^{(0)}}{\sqrt{\lambda_{k}^{(0)}}}\cdot\mathbf{E}_{i}(x,y,z_{2}),\ 
\widetilde{\mathbf{H}}_{k}(x,y,z_{2})=\sum_{i=1}^{N}\frac{v_{k,i}^{(0)}}{\sqrt{\lambda_{k}^{(0)}}}\cdot\mathbf{H}_{i}(x,y,z_{2})\right\}
\end{equation}
in order to fulfill the requirement of unity intensity within the ROI ($m^{(0)}\overset{!}{=}1$, see paragraph ``QME: Spot size operator (QME:SSO)'' in section~\ref{Sec: TheoreticalBackground}). $\mathbf{M}^{(2)}$ is a $N\times{}N$ matrix with the elements given by 
\begin{equation}
\label{eq:s318} 
M_{ij}^{(2)}=\frac{1}{2}\int_{\textrm{ROI}}  |\mathbf{r}-\mathbf{r}_0|^2  \left( \widetilde{\mathbf{E}}_{i}^{\ast}(x,y,z_{2})\times\widetilde{\mathbf{H}}_{j}(x,y,z_{2})+\widetilde{\mathbf{E}}_{i}(x,y,z_{2})\times\widetilde{\mathbf{H}}_{j}^{\ast}(x,y,z_{2})\right) \cdot \mathbf n dS.
\end{equation}
After transforming back to the original base we denote the eigenvalues of ${\mathbf{M}}^{(2)}$ as $\lambda_{k}^{(2)}$ and the eigenvectors as $\mathbf{v}_{k}^{(2)}$. Then, the eigenvector $\mathbf{v}_{\min}^{(2)}$ associated with the smallest eigenvalue $\lambda_{\min}^{(2)}=\min\left(\lambda_{k}^{(2)}\right)$ corresponds to the smallest spot achievable within the ROI. The respective superposition is 
\begin{equation}
\label{eq:s319} \left\{\mathbf{E}_{\min}(x,y,z)=\sum_{i=1}^{N}v_{\min,i}^{(2)}\mathbf{E}_{i}(x,y,z),\ \mathbf{H}_{\min}(x,y,z)=\sum_{i=1}^{N}v_{\min,i}^{(2)}\mathbf{H}_{i}(x,y,z)\right\},
\end{equation}
for $z=z_{1}$ and $z=z_{2}$. We employed this vectorial definition to minimize the size of a focused spot in section~\ref{s:sec34} using a superposition of vector Bessel beams. The spot size minimization performed both in section \ref{s:sec33} using a superposition of LG beams and in the experimental section \ref{s:sec4} is based on the simpler scalar version ot the QME:SSO matrix defined as 
\begin{equation}
\label{eq:s3110}
M_{ij}^{(2)}=\int_{\textrm{ROI}} |\mathbf{r}-\mathbf{r}_{0}|^2 \widetilde{u}_{i}^{\ast}(x,y,z_{2})\widetilde{u}_{j}(x,y,z_{2})dS.
\end{equation} 

\subsection{Maximizing transmission through apertures using Laguerre Gaussian beams}
\label{s:sec32}

\begin{figure}[htb]
\centering\includegraphics[width=13cm]{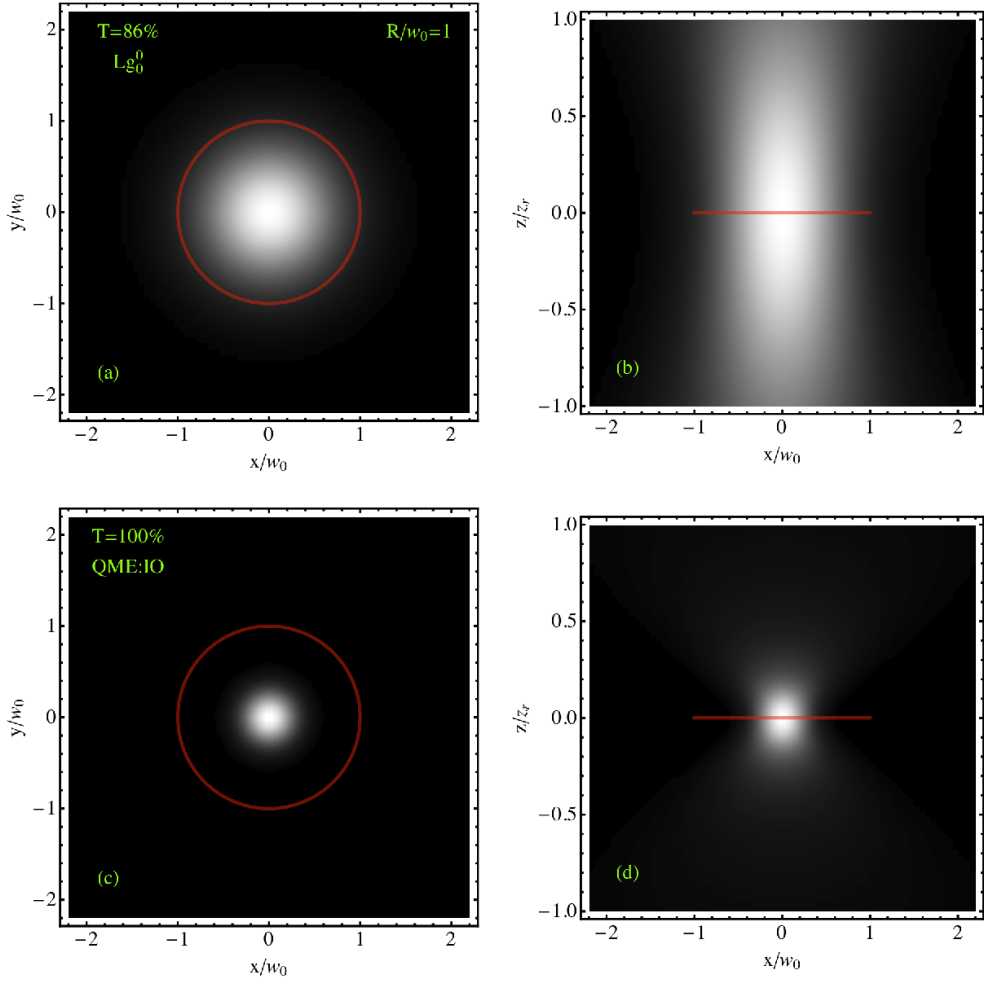}
\caption{2D intensity profiles together with the ROI highlighted by the red circle. The transmittance through the ROI is described by the coefficient $T$. (a) Transversal and (b) longitudinal cross section of a Gaussian beam ($L=0$ and $P=0$). (c) and (d)  QME superposition delivering the highest intensity in the ROI ($R=w_0$).}
\label{fig1}
\end{figure}

\begin{figure}[htb]
\centering\includegraphics[width=13cm]{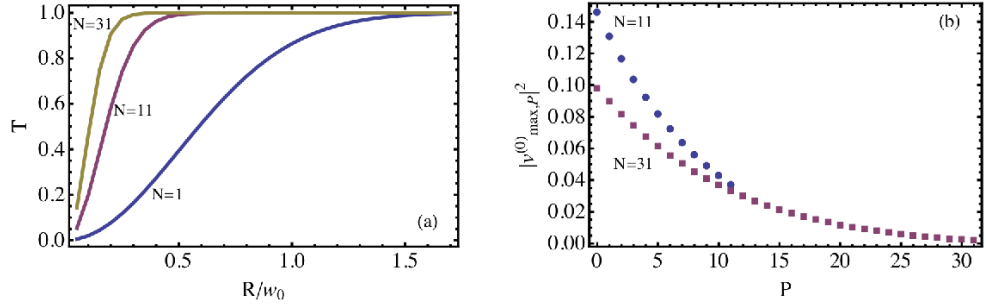}
\caption{(a) Total transmittance through the ROI for the QME intensity optimized beam as a function of the ROI relative radius $R/w_0$ and for different numbers $N$ of LG modes considered. (b) Relative intensity $|v^{(0)}_{\max,P}|^2$ of the LG modes ($P=0..N$) decomposing the QME intensity optimized beam.}
\label{fig2}
\end{figure}

In this subsection, we consider a superposition of LG beams 
propagating in the $z$-direction and modulating the carrier wave $u_c=\exp(-\text{i} (k_0 z-\omega t))$. 
In cylindrical coordinates, LG beams are defined by~\cite{ESiegman:1986p6887}: 
\begin{equation}\label{LGb}
{u}_{P}^L(r,\phi,z)=\frac{\text{i} C_P^L z_r}{q(z)} \left(\frac{\text{i}k_0w_0r}{\sqrt{2}q(z)}\right)^{|L|}
\left(\frac{-q^*(z)}{q(z)}\right)^{P} 
L_P^{|L|} \left(\frac{2r^2}{w^2(z)}\right) 
\exp\left( \frac{-\text{i} k_0r^2}{2q(z)}-\text{i} L \phi\right)
\end{equation}
with $z_r=k_0 w^2_0/2$, $q(z)=z+\text{i} z_r$, and $w^2(z)=1+z^2/z_r^2$ where $w_0$, $k_0$, $\omega$ are the Gaussian beam waist, vacuum wave vector, and optical frequency, respectively. This beam profile is a solution of the paraxial equation for integer values of $P$ and $L$ corresponding respectively to the radial and azimuthal index of the beam. Within the paraxial approximation, the intensity of the beam transmitted through a planar ROI defined by a disk  centered on $r=0$ is proportional to $\int_0^R 2 \pi r |{u}_{P}^{L}(r,\phi,z)|^2 dr$ where $R$ is the radius of the disk.
The coefficient  $C_P^L=\sqrt{{P!}/{(P+|L|)!}}$ is the normalization factor such that the total intensity of the beam, for an infinite ROI, is unity. 

The transmission is maximized using the practical implementation of the QME concept described above in section~\ref{s:sec31}; that is the QME:IO was assembled according to Eq.~\eqref{eq:s315} using the respective representation in cylindrical coordinates. We only considered the radial family of LG beams ($L=0$) and performed the optimization in the plane at $z=0$. Figure~\ref{fig1} shows the final superposition $u_{\max}^{0}(x,y,0)=\sum_{P=1}^{N}v_{\max,P}^{(0)}u_{P}^{0}(x,y,0)$ in the case of a ROI with a radius equal to the waist of the Gaussian envelope. 

In Fig.~\ref{fig2}, we observe that the maximal transmission achievable via the QME intensity optimized beams  depends on the number of LG beams considered in the superposition and on the size of the ROI. Indeed, the QME for a smaller ROI needs a larger number of LG modes to achieve 100\% transmission. 

\subsection{Smallest focal spot using Laguerre Gaussian beams}
\label{s:sec33}

\begin{figure}[htb]
\centering\includegraphics[width=13cm]{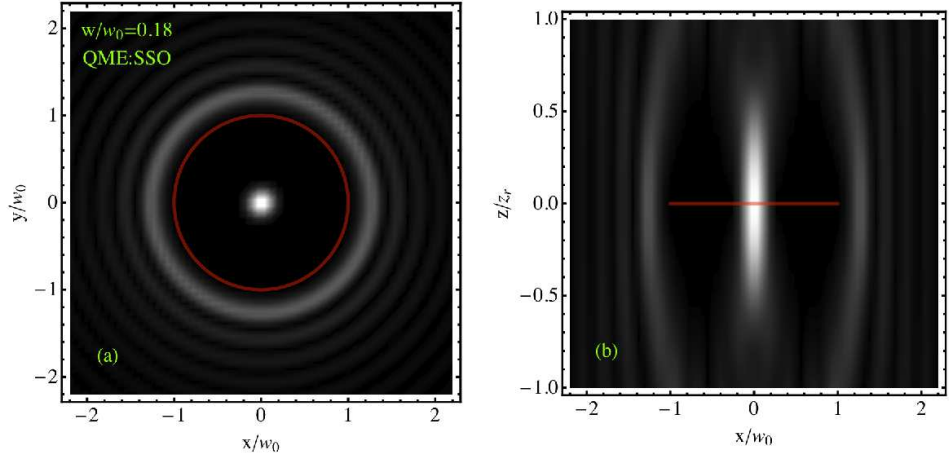}
\caption{(a) Transversal and (b) longitudinal 2D intensity cross sections of the QME superposition delivering the smallest focal spot in the ROI ($R=\lambda$) considering 25 LG modes. $w/w_0$ is the relative SOIM measured according to Eq.~\eqref{Eq: SOM}. The Strehl ratio in (a) is $4.5\%$.}
\label{fig3}
\end{figure}

\begin{figure}[htb]
\centering\includegraphics[width=13cm]{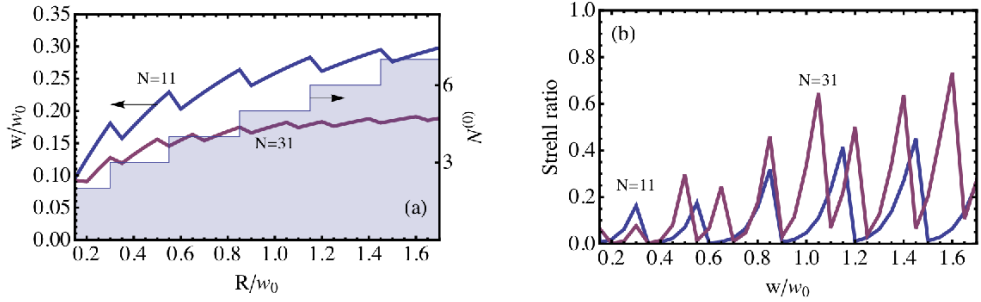}
\caption{(a) Spot size as a function of the radius of the ROI for different number of LG modes considered. The right hand scale and filled curve indicate the numbers of intensity eigenmodes $N^{(0)}$ fulfilling the intensity criteria for the $N=11$ case. The arrows indicate the corresponding scales. (b) Ratio between the ROI intensity of the smallest spot size eigenmode and the largest intensity achievable in the ROI (Strehl ratio).}
\label{fig4}
\end{figure}

Using a superposition of LG beams we have also minimized the size of a focal spot using the representation of the QME:SSO \eqref{eq:s3110} in cylindrical coordinates. It is important to note at this point that we only retain the intensity eigenmodes whose eigenvalues are within a chosen fraction of total intensity. This is equivalent to considering only beams that have a significant intensity contribution in the ROI. Intuitively, the optimization procedure may be performing so well that a spot of size zero is finally obtained if no intensity threshold is applied. Figure~\ref{fig3} shows the smallest spot superposition where we observe the appearance of sidebands just outside the ROI.  These sidebands are a secondary effect of squeezing the light below its diffraction limit. It is these sidebands that decrease the efficiency of the squeezed spot with respect to the maximal possible intensity in the ROI as calculated via the QME:IO. Using the ratio between these two intensities we can define the intensity Strehl ratio~\cite{Sales:1997p9891} for the QME:SSO (see Fig.~\ref{fig4}b). We remark that both, the spot size and the Strehl ratio, show resonances as a function of the ROI size. This can be explained considering the number of intensity eigenmodes used for the spot size operator. Indeed, as the ROI size decreases, so does the number of significant intensity eigenmodes. Each time one of these modes disappears (step in Fig.~\ref{fig4}), we have a sudden increase in the minimum spot size achievable accompanied with an enhanced Strehl ratio as we drop the most intensity inefficient mode. Overall, the Strehl ratios determined in our studies predominantly exceeded values of $1\%$ even when spots were tightly squeezed. Therefore, the observed decrease of intensity is not to severe in terms of potential applications of squeezed beams for optical manipulation and imaging. 

\begin{figure}[htb]
\centering\includegraphics[width=12cm]{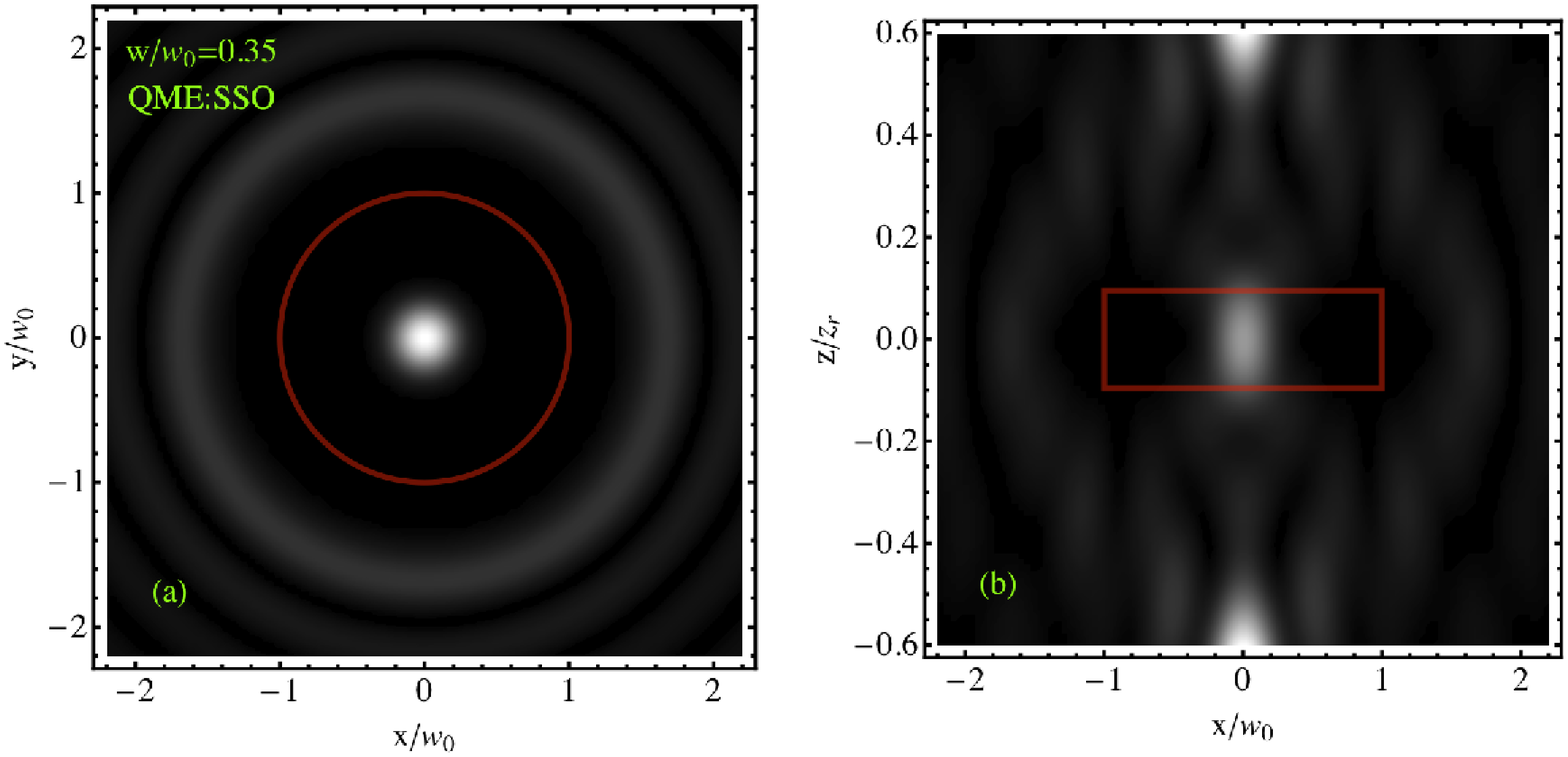}
\centering\includegraphics[width=12cm]{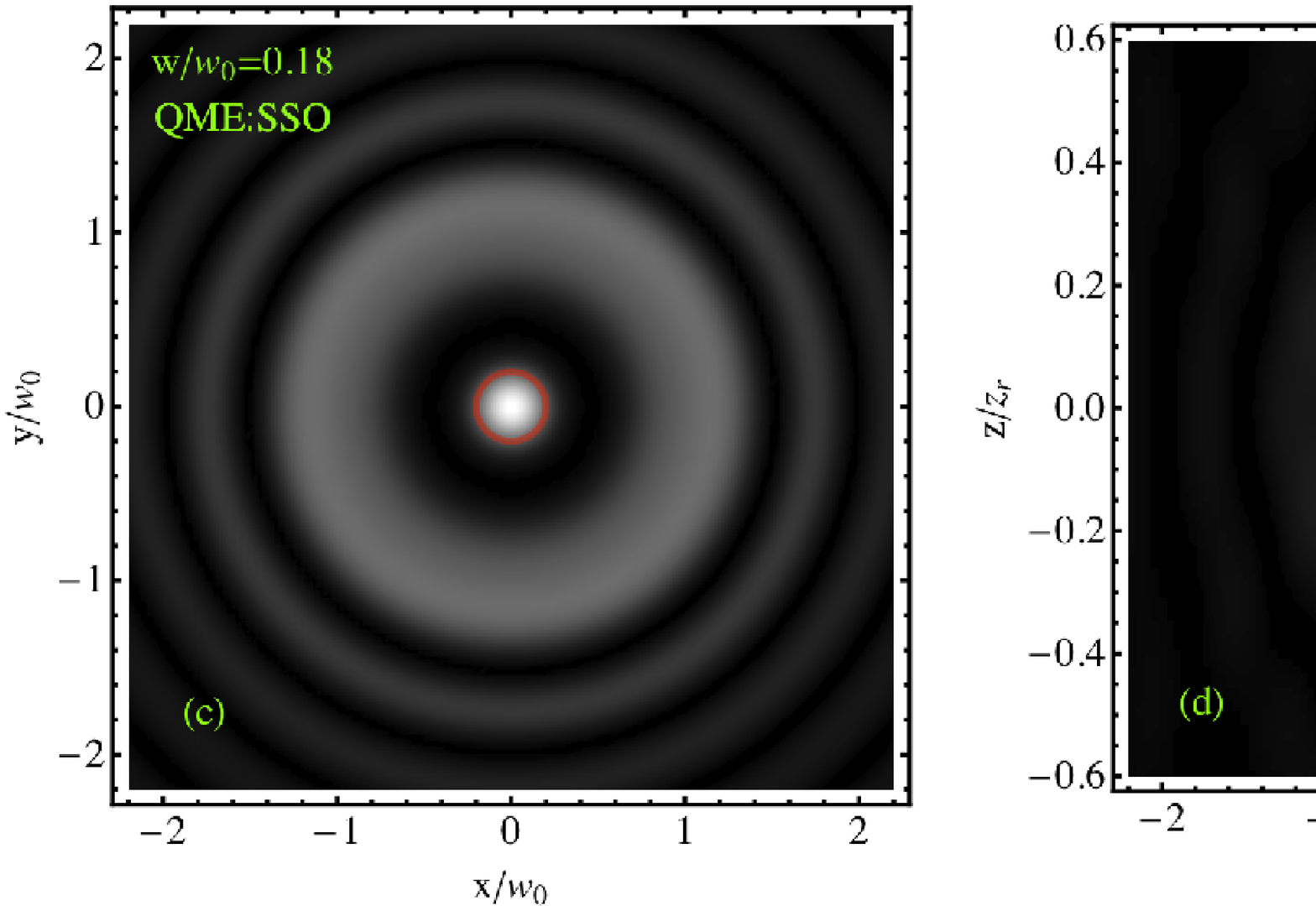}
\caption{(a,c) Transversal and (b,d) longitudinal intensity cross sections of the QME:SSO for two different 3D ROI. Strehl ratio: (a) $5.2\%$, (c) $1.5\%$.}
\label{fig5}
\end{figure}

Analogously to the approach to calculate the smallest spot size in a planar ROI, we can determine the QME:SSO for a volume. In this case, the sidebands appear in the outside of the ROI  in both the lateral and longitudinal directions. The different volumes considered in Fig.~\ref{fig5} suggest that there is an intrinsic link between squeezing light in the lateral direction and in the longitudinal direction. 

\begin{figure}[htb]
\centering\includegraphics[width=12cm]{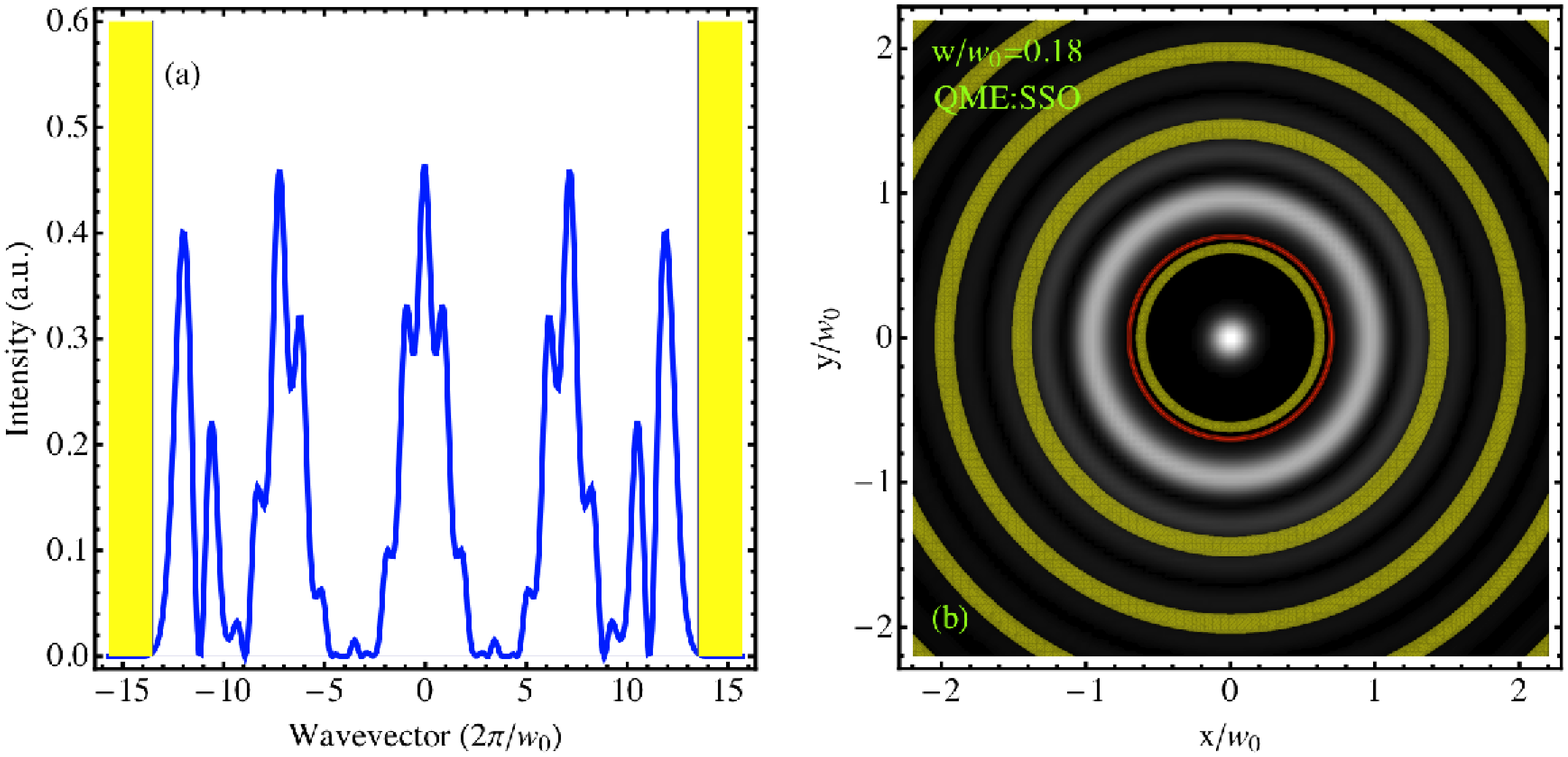}
\caption{(a) Radial wavevector spectral density. Yellow highlights regions outside the spectral bandwidth. (b) Transversal cross section of the QME spot size optimized field intensity with yellow showing super-oscillating regions.}
\label{fig6}
\end{figure}

On a final note, we remark that squeezing light below its diffraction limit may be associated with the effect of super-oscillations~\cite{Berry:2006p2578}. This refers specifically to the ability to have a local $k$-vector (gradient of the phase) larger than the spectral bandwidth of the original field. To visualize this effect, in the case of QME spot size optimized beams, we have calculated the spectral density of the radial wave-vector for the smallest planar spot~\cite{Dennis:12p9892}. As shown in Fig.~\ref{fig6}, this spectral density clearly identifies a spectral bandwidth (white background in Fig.~\ref{fig6}). Regions of the beam which exhibit locally larger wave-vectors than the ones supported by this spectral band width correspond to super-oscillating regions. The local wave vector is defined as $\partial_r \arg(u(r))$ where $\arg(u)$ defines the phase of the analytical  signal $u$. We observe, that super-oscillations occur only in the dark region of the beam. Additionally, when the ROI is large compared to the Gaussian beam waist $w_0$, there are no super-oscillating regions. These only appear when the beam starts to be squeezed.

\subsection{Smallest focal spot using Bessel beams}
\label{s:sec34}

\begin{figure}[htb]
\centering\includegraphics[width=12cm]{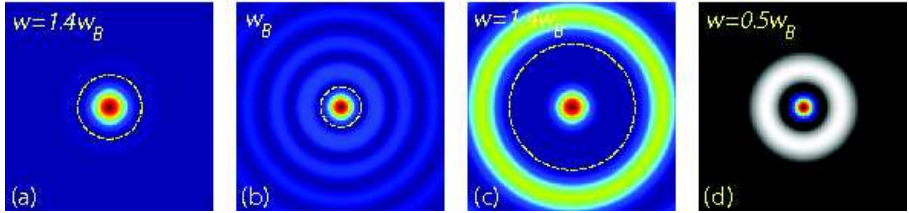}
\caption{Intensity cross sections: (a) Airy disk for the maximum numerical aperture considered NA$=\sin(\theta_{max})=0.1$. The yellow dashed circle shows the position of the smallest zero-intensity circle taken as the ROI inside which the spot size is calculated. The spot size is normalized to the spot size of the reference Bessel beam. (b) Reference Bessel beam corresponding to the largest cone angle $\theta_{max}$. The SOIM of the reference Bessel beam is denoted as $w_{\textrm{B}}$. (c) QME spot size optimized beam for a superposition of Bessel beams ($\theta\in[0,\theta_{max}]$) for a large ROI highlighted by the dashed yellow circle. Strehl ratio: $2\%$. (d) QME spot size optimized beam for a small ROI. Strehl ratio: 0.$2\%$. The gray-scaled region shows the sidebands while the color range the ROI. Notice that the two scales are different.}
\label{fig7}
\end{figure}

\begin{figure}[htb]
\centering\includegraphics[width=12cm]{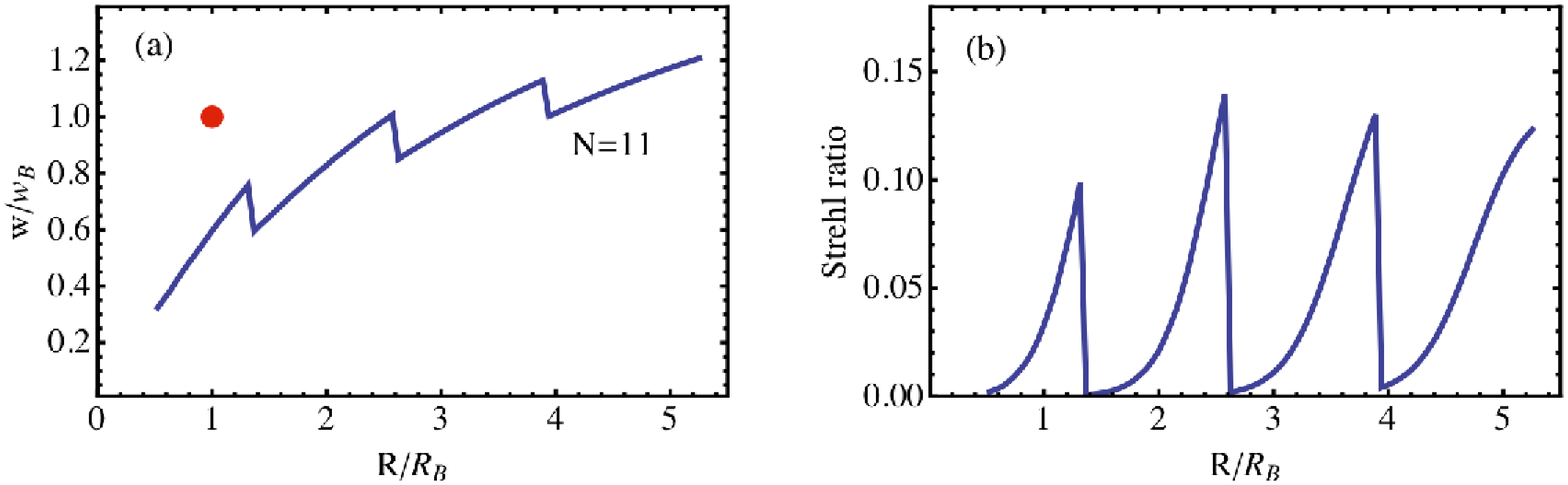}
\caption{(a) Relative spot size $\Delta r/w_{\textrm{B}}$ of the Bessel beam superposition as a function of the relative ROI radius $R/R_{\textrm{B}}$. The SOIM $w_{\textrm{B}}$ and the ROI radius $R_{\textrm{B}}$ are associated with the reference Bessel beam shown in Fig.~\ref{fig7}(b), where the ROI is indicated as dashed circle. For comparison, the red dot indicates the location of the reference beam in the $\Delta r/w_{\textrm{B}}$ vs. $R/R_{\textrm{B}}$ plot. (b) Strehl ratio vs relative ROI radius $R/R_{\textrm{B}}$.}
\label{fig8}
\end{figure}

The paraxial approximation employed above in the case of LG beams can be used to describe sub-diffracting beams but breaks down when beams are tightly focused. As a consequence we must consider full vectorial solutions of Maxwell's equations. Here, we have chosen Bessel beams as a base-set and determined the superposition of Bessel beams which minimized the spot size in a planar finite ROI. Note that the problem of the finite intensity of Bessel beams~\cite{Durnin:1987p7503} is easily circumvented here due to the finite ROI size considered. The monochromatic electric vector field of the vectorial Bessel beam may explicitly be expressed as~\cite{Volke-Sepulveda:2002hk}
\begin{eqnarray}
\label{eq:scal-BB}
\nonumber
\mathbf{E}& =&E_0  \exp\left(iL\phi+ik_t z)\right) \Bigg((\alpha \mathbf{{e}}_x+\beta \mathbf{{e}}_y) J_L(k_t r) \\
\label{eq:s341}
&&+\frac{ik_t}{2k_z}((\alpha+i\beta)\exp(-i\phi)J_{L-1}(k_t r)-(\alpha-i\beta)\exp(i\phi)J_{L+1}(k_t r))\mathbf{{e}}_z\Bigg)
\end{eqnarray}
where $k_t=k_0\sin(\theta) $ and $k_z=k_0\cos(\theta) $ are the transversal and longitudinal wave vectors with $\theta$ the characteristic cone angle of the Bessel beam. $\mathbf{{e}}_{x}$, $\mathbf{{e}}_{y}$ and $\mathbf{{e}}_{z}$ are the unit vectors in the Cartesian coordinate system. The parameter $L$ corresponds to the azimuthal topological charge of the beam while $\alpha$ and $\beta$ are associated with the polarization state of the beam. The magnetic field  $\mathbf{H}$ was deduced according to Maxwell's equations and the QME:IO and QME:SSO operators are assembled according to the expressions \eqref{eq:s313} and \eqref{eq:s318}, respectively. Figure~\ref{fig7} shows a comparison between Airy disk, Bessel beam, and QME optimized spot considering a numerical aperture of NA$=0.1$. As in the case of the LG beams, squeezing the focal spot  is accompanied by side bands and a loss in efficiency shown by the Strehl ratio (see Fig.~\ref{fig8}).

\section{Experimental QME}
\label{s:sec4}

\subsection{Experimental implementation of the QME concept}

\begin{figure}[htb]
\centering\includegraphics[width=0.9\textwidth]{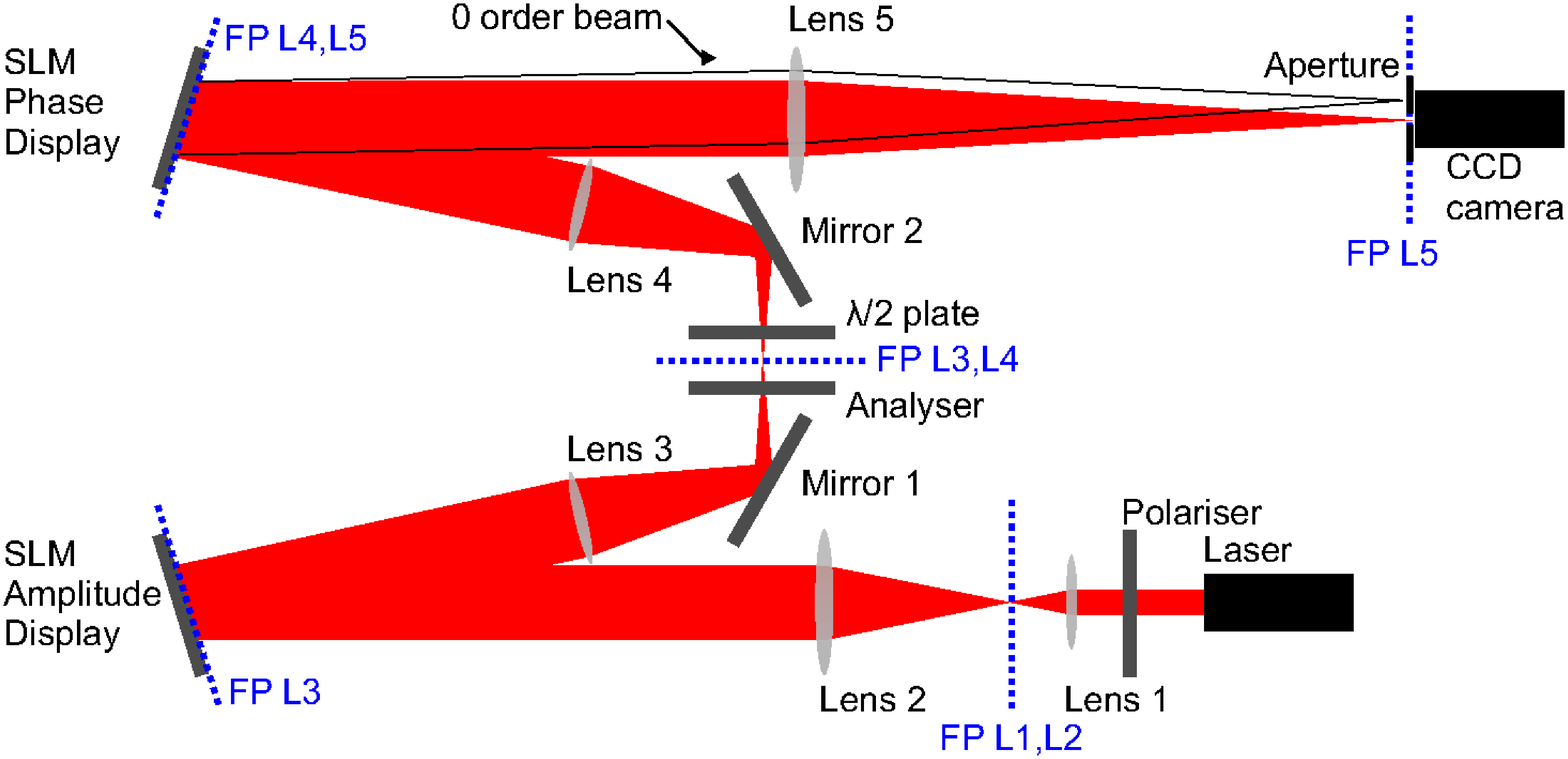}
\caption{Experimental setup. FP = focal plane, L = Lens. Focal widths: $f_{1}=50~\textrm{mm}$, $f_{2}=500~\textrm{mm}$, $f_{3}=f_{4}=400~\textrm{mm}$, $f_{5}=1~\textrm{m}$. Laser: JDS Uniphase HeNe laser, $P_{\max}=10~\textrm{mW}$, $\lambda=633~\textrm{nm}$, SLM: Holoeye HEO 1080 P dual display system, $\textrm{resolution}=1920~\textrm{pixel}\times1080~\textrm{pixel}$, $\textrm{display size}=1~\textrm{in}\times0.7~\textrm{in}$. CCD camera: Basler pilot piA640-210gm, $\textrm{resolution}=648~\textrm{pixel}\times488~\textrm{pixel}$, $\textrm{pixel size}=7.4~\upmu\textrm{m}\times7.4~\upmu\textrm{m}$.}
\label{fig:setup}
\end{figure}

According to the theoretical foundations of the QME concept, the successful experimental implementation within the field of optics requires a \emph{linear} optical system along with the ability to shape laser fields in both amplitude and phase. We have achieved this by using the experimental setup shown in Fig.~\ref{fig:setup}. A $\textrm{HeNe}$ laser beam is expanded and subsequently amplitude modulated by a spatial light modulator (SLM) display operating in conjunction with a pair of crossed polarizers. Analogously to a liquid crystal display on a computer or laptop monitor, the liquid crystal SLM display rotates the polarization of the incident light by an angle depending upon the voltage applied to the display pixels. The amplitude modulated beam is then imaged onto a second SLM display through a pair of lenses. This second SLM display along with a subsequent Fourier lens and aperture served to modulate the phase of the laser beam in the standard first order configuration~\cite{DiLeonardo2007}. The field modulations of interest were encoded as RGB images where the blue channel represented the amplitude and the green channel the phase modulation. The SLM controller extracted these information and applied the two channels to the respective panel. We have performed calibration measurements to ensure that both the amplitude and phase modulation exhibited a linear dependence on the applied 8-bit color value between 0 and 255. A CCD camera allowed us to record images of laser fields in the Fourier plane of lens 5. 

To conform this experimental section to the conventions introduced in section~\ref{s:sec31} we remark that we shaped a set of test fields both in amplitude and phase in the initial plane at $z=z_{1}$ which coincided with the two SLM panels and subsequently minimized the size of a focal spot in the target plane at $z=z_{2}$ which coincided with the CCD camera chip. For our proof-of-principle experiments we ignored polarization effects and considered a set of scalar fields $E_{i}(x,y,z)=A_{i}e^{\textrm{i}\phi_{i}(x,y,z)}$ ($i=1\dots{}N$) where the CCD camera detected the associated intensity $I_{i}(x,y,z_{2})\propto|E_{i}(x,y,z_{2})|^{2}$. Both the QME:IO and the QME:SSO were assembled from these fields according to the scalar expressions \eqref{eq:s315} and \eqref{eq:s3110}, respectively. 

The amplitudes $A_{i}(x,y,z_{2})$ were determined by simply recording an associated intensity image $I_{i}(x,y,z_{2})$ and subsequently taking the square root. We used the three-step phase retrieval algorithm described in Ref.~\cite{Malacara1992OpticalShopTesting} to retrieve the phase modulations $\phi_{i}(x,y,z_{2})$. This algorithm is based on interference with a reference field $E_{\textrm{R}}(x,y,z)=A_{\textrm{R}}(x,y,z)e^{\textrm{i}\phi_{\textrm{R}}(x,y,z)}$. The reference field's intensity was distributed uniformly over the ROIs considered in our experiments by adding a square phase $\phi_{\textrm{R}}\propto(x^{2}+y^{2})$ to the reference field using the SLM phase panel. Moreover, a constant phase term $\phi_{k}=2\pi k/3$ was added for $k=0,1,2$, and the superimposed fields $E_{i}(x,y,z_{1})+E_{\textrm{R}}(x,y,z_{1})e^{\textrm{i}\phi_{k}}$ were then encoded onto the SLM. The associated intensity distributions explicitly read
\begin{align}
 \nonumber
 I_{i,k} &= \left| A_i \right|^2 + \left| A_\text{R} \right|^2 + 2\,\sqrt{A_i\,A_\text{R}} \,\cos \left(\phi_\text{R} - \phi_i + 2\pi\,k/3 \right)
 \\
 \label{eq:s411}
 &=: \mspace{29mu} I_{\text{bg},i} \mspace{29mu} +  \mspace{29mu} \gamma_i \mspace{29mu} \cos \left( \mspace{14mu} \Delta\phi_i \mspace{13mu} + 2\pi\,k/3 \right),
\end{align}
where the spatial coordinates $(x,y,z)$ were omitted for brevity. These three intensity distributions represent a 3-dimensional equation system with three unknowns $I_{\text{bg},i}$, $\gamma_i$ and $\Delta\phi_i$. The latter is explicitly obtained as
\begin{align}
 \label{eq:s412}
 \Delta\phi_i \text{ mod } 2\pi = \text{atan2} \left\{ \left( 1 - \cos \left( 2\pi\,k/3 \right) \right) \left( I_{i,0} - I_{i,2} \right), \, \sin \left( 2\pi\,k/3 \right) \left( 2\,I_{i,1} - I_{i,0} - I_{i,2} \right) \right\},
\end{align}
where $\text{atan2}\{\zeta,\xi\}$ is the two argument arctangent function corresponding to the argument of the complex number $\xi+\textrm{i}\zeta$. Note that the reference phase $\phi_{R}$ cancels out when calculating the operator elements due to multiplication of a complex conjugate field $\propto e^{-\text{i}\phi_{\textrm{R}}(x,y,z)}$ with a complex field $\propto e^{\text{i}\phi_{\textrm{R}}(x,y,z)}$. Therefore Eq.~\eqref{eq:s412} yields the adequate phase modulation required to assemble the QME operators.

During the course of our experiments we verified the linearity of our optical system by performing a comparison between what we term the ``experimental superposition (Exp-S)'' and the ``numerical superposition (Num-S)''. The Exp-S refers to the case where the set of QME optimized  superposition coefficients $a_{i}$ is used to encode the optimized superimposed field onto the SLM. The CCD camera then detected the intensity $I_{\textrm{Exp-S}}(x,y,z_{2})$ corresponding to this encoded optimized field. The Num-S utilizes the fields $E_{i}(x,y,z_{2})$, which were individually measured to assemble the QME operators, in order to \emph{numerically} determine the intensity distribution as $I_{\textrm{Num-S}}(x,y,z_{2})\propto\left|\sum_{i=1}^{N}a_{i}E_{i}(x,y,z_{2})\right|^{2}$. Crucially, linearity is verified if $I_{\textrm{Exp-S}}(x,y,z_{2})=I_{\textrm{Num-S}}(x,y,z_{2})$. This is indeed observed in our experiments as demonstrated in the following subsection which features a comparison of  experimental and numerical intensity distributions. 

\subsection{Results and discussion}

\begin{figure}[htb]
\centering\includegraphics[width=0.8\textwidth]{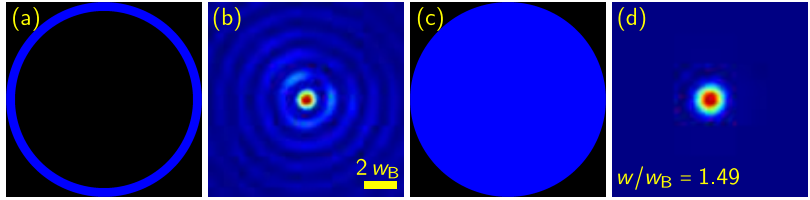}
\caption{SLM encoded field modulations and resulting beam profiles. (a) Ring mask RGB image as encoded onto the dual panel SLM. (b) Associated Bessel beam created in the CCD camera plane. (c) Aperture RGB image as encoded onto the dual panel SLM. (d) Associated Airy disk as detected by the CCD camera. The yellow bar in (b) represents 2 times the SOIM $w_\text{B}$ of the Bessel beam's central core. $w$ in (d) is the SOIM of the Airy disk.}
\label{fig:bessels}
\end{figure}

In our experiments, we used $N = 11$ non overlapping amplitude ring masks with a constant phase modulation as fields of interest $E_{i}(x,y,z_{1})$. After propagation through the Fourier lens 5 (see Fig.~\ref{fig:setup}) the resulting fields $E_{i}(x,y,z_{2})$ form a set of Bessel beams. Figure~\ref{fig:bessels}(a) shows the largest ring modulation encoded onto the SLM with the resulting Bessel beam shown in Fig.~(b). As this particular Bessel beam comes along with the highest NA compared to the Bessel beams created with smaller ring modulations, the beam shown in 
Fig. (b) exhibits the smallest central spot of all beams realized in our experiments. The SOIM of the Bessel beam featuring the smallest core is denoted as $w_\text{B}$ and used as reference for the measurements presented below. For comparison Figure~\ref{fig:bessels}(c) depicts a circular aperture which is encoded onto the SLM in order to observe the Airy disk (see Fig.~(d)). The SOIM of the Airy disk is approximately $1.5$ times larger than the core of the reference Bessel beam as expected~\cite{Wang:2008p9893}. 

\begin{figure}[htb]
\centering\includegraphics[width=0.8\textwidth]{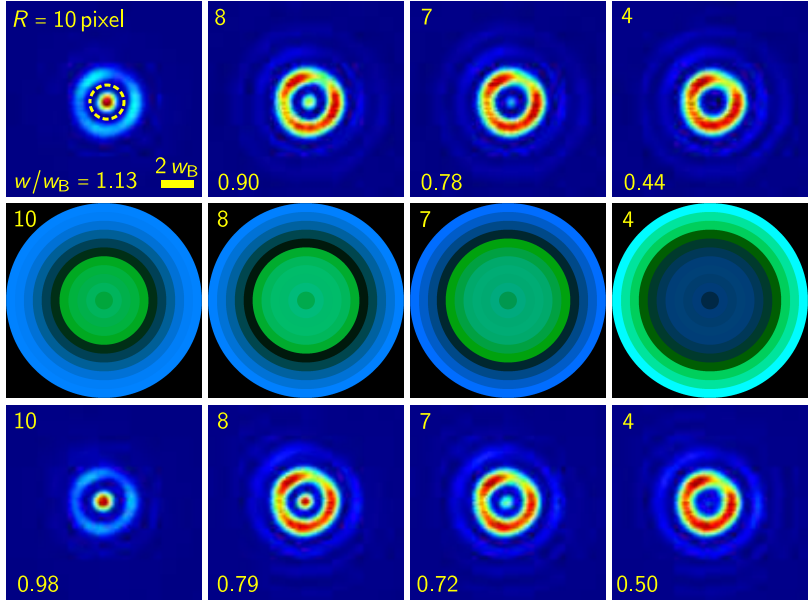}
\caption{Experimental QME spot size minimization. \emph{Top row}: $I_{\textrm{Num-S}}(x,y,z_{2})$ for different ROI radii in pixel as indicated in the top left corner of all graphs shown. The ROI is exemplary indicated as a dashed ring in the left hand side intensity distribution. The number in the bottom left corner represents the SOIM $w$ in units of the reference SOIM $w_{\textrm{B}}$. \emph{Central row}: Optimized experimental distribution as RGB encoded onto the SLM. \emph{Bottom row}: Intensity distributions $I_{\textrm{Exp-S}}(x,y,z_{2})$. The relative SOIM $w/w_\text{B}$ is indicated in the lower left corner.}
\label{fig:spots}
\end{figure}

The results of the performed QME spot size minimization are shown in Fig.~\ref{fig:spots} for different sizes of the ROI. To begin with, the comparison of the Num-S intensity distribution $I_{\textrm{Num-S}}(x,y,z_{2})$ (top row) and the Exp-S intensity distributions $I_{\textrm{Exp-S}}(x,y,z_{2})$ (bottom row) clearly reveals good agreement and thus verifies the linearity of our optical system as elucidated above. For completeness, the central row shows the Exp-S superposition in RGB format as encoded onto the SLM. The color code features a blue channel representing the amplitude modulation from 0 (black) to 1 (blue) and a green channel corresponding to phase modulations from 0 (black) to $2\pi$ (green). Next, we conclude from the measured relative SOIM $w/w_\text{B}$ that the spot size decreases if the ROI size is reduced. The reduced spot size is achieved at the expense of the spot intensity which is redistributed to a ring outside of the ROI similar to the theoretical results presented in section~\ref{s:sec34} and Fig.~\ref{fig7}. Referring to the Exp-S data, for $R = 7\,\text{pixel}$ the spot size is reduced to $72\,\%$ of the size of the reference Bessel beam's core and even further to $50\,\%$ for $R = 4\,\text{pixel}$. The latter result is somewhat vague, though, due to the low spot intensity which may be truncated by the sensitivity threshold of the CCD detector and thus may appear smaller. However, our experimental results overall clearly verify the QME concept applied to spot size minimization. Moreover, the results strongly suggest that the QME optimization may indeed squeeze spots to the subdiffractive regime since the optimal superposition of Bessel beams not only beats the Airy disk but also the reference Bessel beam. 

\begin{figure}[htb]
\centering\includegraphics[width=0.6\textwidth]{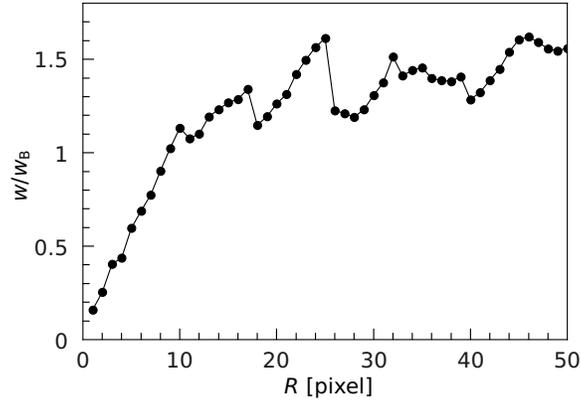}
\caption{Relative SOIM $w/w_\text{B}$ within the ROI depending on the ROI radius $R$ in pixel for numerical superposition of the fields $E_{i}(x,y,z_{2})$ in the CCD plane.}
\label{fig:SOMvsR}
\end{figure}

We have performed an additional comparison of the experimental results to the theoretically predicted ones: The relative SOIM $w/w_\text{B}$ was evaluated for the Num-S  for ROI radii ranging from $R = 1\,\text{pixel}$ to $R = 50\,\text{pixel}$ in steps of one pixel (see Fig.~\ref{fig:SOMvsR}). In comparison to the corresponding graph shown in Fig.~\ref{fig8} we observe agreement in terms of both the general decrease of $w/w_\text{B}$ with decreasing $R$ and the peaks occurring periodically along the $R$-axis. Surprisingly, for $R \geq 10$, we do not observe values of $w/w_\text{B}\approx1$; that is the QME spot minimization does not deliver, as one might naively expect, the reference Bessel beam featuring the smallest SOIM $w_\text{B}$ of all Bessel beams considered. This is due to the ring structure of Bessel beams which significantly adds to the SOIM and therefore to the spot size in the case of large ROIs. As a consequence, the QME optimization aims to superimpose the set of Bessel beams in a manner that the rings are destructively interfered within the ROI, only retaining the central core. Given that the ring structure is essential for the reduced core size of Bessel beams compared to the diffraction limited Airy disk we observe values of $w$ which are larger than $w_{\textrm{B}}$. Indeed, the values of $w$ are fairly close to the size of the Airy disk which was determined as $w\approx1.5w_{\textrm{B}}$. This strongly suggests that for large ROIs the QME spot minimization yields a superposition of Bessel beams which closely resembles the diffraction limited Airy disk. Finally, we remark that for $R < 7$ the relative SOIM $w/w_\text{B}$ becomes very small and has to be taken with care due to the possible truncation of the spot mediated by the CCD detector's sensitivity threshold as mentioned above. 

\section{Discussion and Conclusion}

We have theoretically and experimentally demonstrated a novel approach based on quadratic measures eigenmodes that enables the optimization of different optical measures. The theory that we employ is rigorous and based on considering the light-matter interaction as a quadratic measure originating from the fields described by Maxwell's equations. Excitingly, we can define many quadratic measure operators to which our approach is applicable (see Table \ref{tab}). The method is thus very powerful and the generic nature of our approach means that it may be applied for example to optimize the size and contrast of optical dark vortices, the Raman scattering or fluorescence of any samples, the optical dipole force, and the angular/linear momentum transfer in optical manipulation. In the present paper we have verified the rigor of the method by demonstrating experimental spot size operator and intensity operator optimization using Laguerre-Gaussian and Bessel light modes using a dual SLM approach to implement the technique. We envisage the QME approach as providing a powerful and versatile theoretical and practical toolbox. Our generic approach is applicable to all linear physical phenomena where generalized fields interfere to give rise to quadratic measures.

\section*{Acknowledgements}
We thank the EPSRC Nanoscope Basic Technology consortium for funding. Mark Dennis is acknowledged for the introduction to super-oscillations. KD is a Royal Society-Wolfson Merit Award Holder. 
\end{document}